\documentclass[prd,onecolumn,10pt,amsfonts]{revtex4}
\usepackage{graphicx} 
\usepackage[latin1]{inputenc} 

\newcommand{\sref}[1]{section \ref{#1}}
\newcommand{\eref}[1]{(\ref{#1})}
\newcommand{\fref}[1]{figure \ref{#1}}
\newcommand{\Fref}[1]{Figure \ref{#1}}
\newcommand{\Eref}[1]{Equation (\ref{#1})}

\begin{document}

\title{On the evolution of primordial gravitational waves:\\ a semi-analytic detailed approach}

\author{M. Soares-Santos}
\thanks{e-mail: msoares@astro.iag.usp.br}
\affiliation{Instituto de Astronomia, Geofísica e Ciências Atmosféricas, 
Universidade de São Paulo, Rua do Matão 1226, Cidade Universitária 05508-090,
São Paulo SP Brazil} 

\author{E. M. de Gouveia Dal Pino}
\thanks{e-mail: dalpino@astro.iag.usp.br}
\affiliation{Instituto de Astronomia, Geofísica e Ciências Atmosféricas, 
Universidade de São Paulo, Rua do Matão 1226, Cidade Universitária 05508-090,
São Paulo SP Brazil}

\date{\today}

\begin{abstract}
A cosmological  gravitational wave background resulting from space-time
quantum perturbations at energy scales  of $\sim 10^{15}$GeV is expected  as
a %straightforward
consequence of the general relativity theory in the context
of the standard cosmological model.
Initial conditions %for the problem
are determined during the inflationary (de Sitter) era, at $z
\gtrsim 10^{25} $. A semi-analytic method was developed to evolve
the system up to the present with no need of simplifying
approximations as the thin-horizon (super-adiabatic) or  the
instantaneous transitions between the successive phases of domain
of the different cosmic fluids. The accuracy of such assumptions,
broadly employed in the literature, is put in check. Since the
physical nature of the fluid (known as dark energy) leading to the
accelerated expansion observed in the recent Universe  is still
uncertain, four categories of models were analyzed: cosmological
constant, X-fluid (phantom or not), generalized Chaplygin gas and
(a parametric form of) quintessence. The results are conclusive
with respect to the insensitivity of gravitational waves to  dark
energy, due to the
%fact that its phase of domain is recent ($z \sim 1 $).
recentness of its phase of domain  ($z \sim 1 $). The empirical
counterparts of the gravitational wave forecasts are still
nonexistent for the noise levels and operational frequencies of
the experiments already built are inadequate to detect those
relics. Perspectives are more promising for  space detectors
(planned to be sensitive to amplitudes of $\sim 10^{-23}$ at
$10^{-3}-1$Hz).
%, but those should not enter in operation before 2015.
The cosmic microwave background is also discussed as an alternative of
indirect detection and the energy density scale of inflation is constrained to
be smaller than $10^{-10}$ in the analysis here presented.
\end{abstract}

\pacs{04.30.-w,98.80.-k} 

\maketitle

%%%%%%%%%%%%%%%%%%%%%%%%%%%%%%%%%%%%%%%%%%%%
%% MAINMATTER
%%%%%%%%%%%%%%%%%%%%%%%%%%%%%%%%%%%%%%%%%%%%

\section{Introduction}\label{intro}
The progresses accomplished in both experimental and observational fields
during the last century permitted to establish a cosmological model in
reasonable agreement with the reality (or at least with the glimpses
provided by the experiments thereof).
One of the most outstanding predictions emerged from that scenario
is the existence of a cosmological background of gravitational waves.
This forecast, almost as old as the general relativity theory itself
(in 1916 Einstein published a work on linearized
weak waves emitted by bodies with null self-gravitation and
propagating through a flat space-time \cite{Einstein:16}),
was indirectly verified only in the seventies (ever since the binary system
PSR 1913+16 has been monitored, showing an orbital deceleration rate
compatible with a kinetic energy dissipation due to gravitational
radiation \cite{Taylor:76,Gullahorn:78}) and up to the present
no direct observational effort succeeded (a somewhat embarrassing fact in the
so-called era of precision cosmology). Nevertheless
it would be hard to believe that this very consequence of the
standard cosmological model (which explains %with relative precision
the thermal history of the Universe, the genesis of large scale
structures and the primordial nucleosynthesis) is incorrect.

There is an immediate analogy between gravitational and electromagnetic
waves. More important are the differences though.
Only large massive objects in movement or coherent space-time vibrations emit
gravitational waves,
while  electromagnetic radiation arises from incoherent
superpositions of individual contributions of atoms, electrons and charged
particles;
near extreme gravitational fields (e.g.,~in the black hole
surroundings) electromagnetic waves tend to darken, while their
gravitational analog tends to be better emitted;
unlike  photons, that interact easily with  matter,  gravitons
(the quanta of gravitational waves)  interact very weakly, being able to
cross regions as dense as the nucleus of a supernova or the primordial
plasma just $10^{-44}$ seconds after the Big Bang.
These differences imply that if (not to say when) gravitational waves
are detected and studied a completely new viewpoint (unaccessible to
electromagnetic-only experiments) will be achieved and the consequences
of that %(even a scientific revolution, according to some enthusiasts)
are unpredictable.

The present work concerns the generation and evolution of
primordial gravitational waves. Quantum perturbations of the
space-time during a de Sitter inflationary phase are computed and
taken as initial conditions for the subsequent evolution, governed
by an excited oscillator-like equation. The evolution of those
primordial perturbations up to the recent Universe is followed
through a semi-analytic method. Since the gravitational wave
background is stochastic (isotropic and stationary) all the
meaningful information is encapsulated in the frequency spectrum
which is analyzed in \sref{gw spectrum}. Other authors
\cite{Grishchuk:05,Zhang:05,Zhang:06,Efstathiou:06} accomplished
similar calculations using analytic approaches which are compared
with the results reported in \sref{decelerated}.

Perspectives of direct detection by both already built and planned
experiments are treated in \sref{direct}. A promising indirect
possibility also discussed in \sref{cmb} is the search for
gravitational wave signatures on the angular spectrum of the
cosmic microwave background (hereafter CMB), which corresponds to
photons that decoupled $\sim 300 000$ years after the Big Bang,
being sensitive to  perturbations present in the cosmic plasma
since that time. The analysis of  possible observational
counterparts characterizes the present work as a theoretical study
able to establish forecasts for future experiments.

About the recent Universe it is remarkable that
the physical nature of the fluid (dark energy) representing
$\sim 70 \% $ of the present total energy density and leading to the
accelerated expansion inferred by supernovae observations remains uncertain.
A plethora of possible dark energy models have been proposed but
the observational constraints are still not enough to determine its
equation of state.
Since dark energy seems to interact only via gravitation, a
reasonable expectation could be to find signatures of its
equation of state on the gravitational wave spectrum. The analysis
for
the most important models
is done in \sref{accelerated}.

%As a last remark, the authors emphasize that the natural unit system
%($\hbar=k_B=c=1 $) is used in the whole text.

\section{Formal developments for primordial tensor perturbations}
\label{formal developments}

In the linear regime, tensor $(t)$ perturbations of a given flat,
homogeneous and isotropic background $(B)$ metric
%\begin{equation}
%ds^{2 \ (B)}=a^2(\tau)\left[d\tau^2-\delta_{ij}dx^idx^j\right]
%\end{equation}
%-- where $\tau$ and $\delta_{ij}$ are the conformal time and the
%Kronecker delta, respectively --
are represented by a transverse-traceless tensor $h_{ij}$ to be
added to the spatial part of the unperturbed metric tensor
\cite{Mukhanov:92}:
\begin{equation}\label{pert metric tensor}
g_{\mu \nu} \equiv g^{(B)}_{\mu \nu}+\delta g^{(t)}_{\mu \nu}=
-a^2(\tau) \left(
\begin{array}{cc}
-1 & 0 \\
0 & \delta_{ij}+h_{ij} \\
\end{array}
\right)
\end{equation}
-- where $\tau$, $a(\tau)$ and $\delta_{ij}$ are the conformal time, the
scale factor and the Kronecker delta, respectively.
In order to compute the evolution of these perturbations the
Einstein-Hilbert action is expanded up to the second order
\cite{Maldacena:03}:
\begin{equation}\label{2nd order action}
S_{\mathrm{O}(2)}=\frac{1}{64 \pi G} \int d^4x \sqrt{g^{(B)}}
\left[-g^{(B)\mu \nu} h_{ij,\mu}h_{ij,\nu} %+ \frac{1}{2}\Pi_{ij}h_{ij}
\right],
\end{equation}
neglecting the anisotropic stress tensor, which would act as
%$\Pi_{ij}=T_{ij}-p\delta_{ij}$ (where $T_{ij}$ is the energy-stress tensor
%and $p$ is the fluid pressure) in the above equation acts as
a source term. % subject to the constraints $\Pi_{ii}=\Pi_{ij,i}=0 $.
The variational principle applied to \eref{2nd order action} implies that
\cite{Mukhanov:92,Giovannini:05}
\begin{equation}\label{h complete}
h''_{ij}+2\frac{a'}{a}h'_{ij}-\nabla^2 h_{ij}=0 %16\pi Ga^2\Pi_{ij}
\end{equation}
(derivatives with respect to $\tau$ being denoted by primes).
One can also write \eref{h complete} in terms  of the cosmic time
$t=\int a(\tau) d\tau$
(with derivatives represented by dots) \cite{JW:W72}: % and,
%neglecting the source term, the resulting equation is
\begin{equation}\label{h(t) w/o source}
\ddot{{\mathsf h}}_{ij} - \frac{\dot{a}}{a}  \dot{{\mathsf h}}_{ij}-
2\frac{\ddot{a}}{a}{\mathsf h}_{ij}-\frac{1}{a^2} \nabla^2 {\mathsf h}_{ij}
=0.
\end{equation}
The conformal perturbation $h_{ij}$ is
related to its cosmic analogous  by
${\mathsf h}_{ij} \equiv a^2(\tau) h_{ij} $ and its
Fourier expansion may be expressed as
\begin{equation}
\label{fouriertransform}
{\mathsf h}_{ij}^{(\lambda)}({\bf x},t)=
\sqrt{16\pi G} \int \frac{d{\bf k}}{(2\pi)^{3/2}} \
\tilde{{\mathsf h}}^{(\lambda)}({\bf k},t) \ e^{(\lambda)}_{ij}({\bf k}) \
\exp\{-i{ \bf k \cdot x}\},
\end{equation}
where $(\lambda)$ accounts for the two polarizations, $(+)$ and $(\times)$.
For a wave propagating in the ${\bf z}$ direction, the polarization  basis
components are:
\begin{equation}\label{basis def}
e^{(\times)}_{ij}=\hat{e}_x \otimes \hat{e}_y + \hat{e}_y \otimes \hat{e}_x ,
\hspace{1cm}
e^{(+)}_{ij}=\hat{e}_x \otimes \hat{e}_x + \hat{e}_y \otimes \hat{e}_y.
\end{equation}
This basis is symmetric, transverse,  traceless and  normalized -- these
properties are, respectively, expressed by the constraints \cite{Boyle:05}
\begin{eqnarray}\label{basis constraints}
e^{(\lambda)}_{ij}({\mathbf k})=e^{(\lambda)}_{ji}({\mathbf k}), \hspace{1cm}
k^i e^{(\lambda)}_{ij}({\mathbf k}) =0, \hspace{1cm}
e^{(\lambda)}_{ii}({\mathbf k}) = 0 \hspace{1cm} \mathrm{and} \nonumber \\
\label{basis constr}
e^{(\lambda)}_{ij}({\mathbf k})e^{(\lambda') \ ij  *}({\mathbf k})=
2\delta_{\lambda \lambda'}, \hspace{1cm}
e^{(\lambda)}_{ij}({\mathbf k})=e^{(\lambda)*}_{ij}(-{\mathbf k}).
\end{eqnarray}
Using \eref{fouriertransform}, one may Fourier-transform
\eref{h(t) w/o source} into
\begin{equation}
\label{gw1}
 \ddot{{\mathsf h}} - \frac{\dot{a}}{a} \dot{{\mathsf h}}
 + \left( \frac{k^2}{a^2} - 2 \frac{\ddot{a}}{a} \right) {\mathsf h} = 0 ,
\hspace{1cm} {\mathsf h} \equiv \tilde{{\mathsf h}}^{(\lambda)}(k,t)
\end{equation}
or into its conformal analogous
\begin{eqnarray}
\label{gw2}
h''+ 2 \frac{a'}{a}h' + k^2 h =0 ,
\hspace{1cm}  h \equiv \tilde{h}^{(\lambda)}(k,\tau) \hspace{1cm}
\mathrm{or \ yet}
\\
\label{gw3}
\mu'' + \left( k^2 - \frac{a''}{a}  \right) \mu =0 ,
\hspace{1cm} \mu \equiv h \ a .
\end{eqnarray}
Equation \eref{gw3} above represents an  oscillator excited by
an effective potential $a''/a$ and is a crucial point of the first
simplifying assumption to be discussed in \sref{assumptions}.

%Neglecting the source term and
Considering once more the Fourier-expanded
perturbation \eref{fouriertransform}, the second order action
\eref{2nd order action} is rewritten in a form which may be promptly
quantized \cite{Boyle:05}:
\begin{equation}\label{action2}
S_{\mathrm{O}(2)}=\sum_{\lambda}\int \frac{a^2}{2}
\left[{h^{(\lambda)}}'{h^{(\lambda)*}}'-k^2h^{(\lambda)}h^{(\lambda)*}
\right]d\tau d{\mathbf k}.
\end{equation}
The conjugate momentum is  $\pi^{(\lambda)}=a^2 {h^{(\lambda)*}}'$,
the corresponding operators are  $\hat{h}^{(\lambda)}_{\mathbf k}$ and
$\hat{\pi}^{(\lambda)}_{\mathbf k}$ and
the commutation relations to be satisfied are
\begin{equation}\label{commutation relations}
\left[\hat{h}^{(\lambda)}_{{\mathbf k}},\hat{h}^{(\lambda')}_{{\mathbf k}'}
\right]=
\left[\hat{\pi}^{(\lambda)}_{{\mathbf k}},\hat{\pi}^{(\lambda')}_{{\mathbf k}'}
\right]=0,
\hspace{1cm}
\left[\hat{h}^{(\lambda)}_{{\mathbf k}},\hat{\pi}^{(\lambda')}_{{\mathbf k}'}
\right]= i\delta^{\lambda \lambda'}\delta({\mathbf k}-{\mathbf k}').
\end{equation}
Since $\hat{h}^{(\lambda)}_{\mathbf k}$ is hermitian its decomposition
in terms of the creation and annihilation operators --  considering the
commutation relations \eref{commutation relations} -- is
\begin{equation}\label{expand h em a}
\hat{h}^{(\lambda)}_{{\mathbf k}}(\tau)=
h[\hat{a}^{(\lambda)}_{{\mathbf k}}]^{(+)}+
h^{*}[\hat{a}^{(\lambda)}_{-{\mathbf k}}]^{(-)},
\end{equation}
where the evolution of $h$ is governed by \eref{gw2} and the
standard commutation relations (for boson particles) are preserved:
\begin{equation}\label{commutation a}
\left[a_k^{-},a_{k'}^{-} \right]=\left[a_k^{+},a_{k'}^{+} \right]=0,
\hspace{1cm} \left[a_k^{-},a_{k'}^{+} \right]=\delta_{kk'}.
\end{equation}

The power spectrum $\Delta^2_t(k,\tau)$ is easily obtained from the above
formalism. Considering equations \eref{expand h em a} and
\eref{fouriertransform} under the constraints imposed by the
 commutation relations \eref{commutation a} and the normalization
properties \eref{basis constr} of
the  transverse-traceless tensor basis  -- and, just for sake of clarity,
expliciting all the physical constants -- the two-point correlation function
is given by
\begin{eqnarray}\label{auto-correlation}
\langle 0|\hat{h}_{ij}(\tau,{\mathbf x})\hat{h}^{ij}(\tau,{\mathbf x'})|0
\rangle & = & \delta({\mathbf x}-{\mathbf x'})
\sum_{\lambda} \frac{V}{(2\pi)^3} \frac{16 \pi G \hbar}{c^3}
\int 2 |h^{(\lambda)}|^2 d{\mathbf k} \nonumber\\
%& = & \frac{V}{(2\pi)^3} \frac{16 \pi G \hbar}{c^3}
%\int 4 |h|^2  4\pi k^3 \frac{dk}{k} \nonumber\\
& = & \delta({\mathbf x}-{\mathbf x'})
\int  \frac{64 \pi G \hbar}{c^3}\frac{k^3}{2\pi^2} |h|^2  d\ln k,
\end{eqnarray}
where $V$ is the volume of the system (normalized to the unity); the
power spectrum is:
% related to the correlation function \eref{auto-correlation} by
\begin{equation}\label{pwr spct t def}
\Delta^2_t(k, \tau) \equiv
\frac{d \langle 0|\hat{h}_{ij}(\tau,{\mathbf x})\hat{h}^{ij}(\tau,{\mathbf x})|0\rangle}{d\ln k}
\end{equation}
and therefore, from \eref{auto-correlation} and \eref{pwr spct t def} and
recovering the natural system of units,
it is straightforward to notice that
\begin{equation}\label{pwr spectrum t}
\Delta^2_t(k, \tau) = 64\pi G \frac{k^3}{2\pi^2}|h|^2.
\end{equation}

\section{An effective equation of state and the scale factor evolution}
\label{effectiveEOS} The standard (Big Bang + inflation)
cosmological model states that \cite{AW:KT90,Coles:05}: (1) in the
very beginning the Universe experiences a de Sitter-like
inflationary phase, when the perturbations are originated; soon
afterwards, (2) the inflationary scalar field decays quickly and
the Universe becomes radiation-dominated ($p_r=1/3\rho_r $); as
the Universe expands, the radiation  density falls as $a^{-4}$ and
at $z \sim 1000 $ (3) the era of cold dark matter (CDM, $p_m=0$)
starts; persisting until $z \sim 1 $, when (4) the expansion
becomes accelerated and the dominant fluid, dark energy, is
characterized by a negative pressure.

The first attempts to explain the dark energy tried to associate that
negative pressure with the vacuum energy ($\Lambda$). The concordance model
in that context would be a Big Bang + inflation + $\Lambda$CDM model.
However, the concordance model faces great difficulties
(the cosmological constant problem) and several alternative
models have been proposed (see \cite{Copeland:06,Lima:04} for review, and
references therein).
Here, four models were chosen for a comparative analysis:
cosmological constant -- the paradigmatic model, with an equation
of state $p=-\rho$;
X-fluid -- a generalization over the $\Lambda$
equation of state,  $p=\omega \rho, \ \omega <0 $, which is
known as phantom fluid
when $\omega <-1 $ \cite{Caldwell:02};
a first order parametrization ($\omega (z) = \omega_0 + \omega_1 z $)
of quintessence,  which is a class of models where a
scalar field is invoked to derive an evolving equation of state;
and generalized Chaplygin gas, with an equation of state
$p=-A/\rho^{\alpha}=-\bar{A}\rho_0 (\rho_0/\rho)^{\alpha} $, $\bar{A},\alpha
\in [0,1] $, that behaves as matter for high $z$ and as
$\Lambda$ in recent eras.

Given the equation of state, the energy density $\rho(x)$
and  pressure $p(x)$ evolutions are easily computed:
\begin{equation}
\rho(x) = \sum_{i=r,m,de} \rho_{i_0} \Psi_i (x), \hspace{1cm}
p(x)    = \sum_{i=r,m,de} \rho_{i_0} \Phi_i (x),
\end{equation}
where $x = 1+z = 1/a$ and  the sum contains three terms: radiation ($r$),
matter ($m$) and dark energy ($de$), this last one corresponding to one of the
models mentioned above. The parametrization in terms
of the functions $\Phi_i(x)$ and $\Psi_i(x)$ is
convenient to simplify the notation.
With this parametrization, the Friedmann and Raychaudhuri equations assume,
respectively, the form
\begin{equation}
\label{friedmann}
\frac{\dot{a}}{a}   =  H_0 \sum_{i} \Omega_{i_0} \Psi_i(x), \hspace{1cm}
\frac{\ddot{a}}{a}  =  -\frac{H_0^2}{2} \sum_{i}
\Omega_{i_0} \left[ \Psi_i(x) +3 \Phi_i(x) \right],
\end{equation}
where $H_0$ and $\Omega_{i_0}$ are the present values of the Hubble
($H=\ddot{a}/a$) and density ($\Omega_i=3H^2\rho_i/8\pi G$) parameters.
The functions $\Phi_i(x)$ and $\Psi_i(x)$ may be written as:
\begin{equation}\label{PsiPhi}%\fl
\begin{array}{lcl}
\Psi_r(x)=  x^{4}, & \hspace{1cm} & \Phi_r(x)=  x^4/3,\\
\Psi_m(x)=  x^{3}, & \hspace{1cm} & \Phi_m(x)=  0,\\
\Psi_{\Lambda}(x)= 1, & \hspace{1cm} &
                \Phi_{\Lambda}(x)=-1, \\
\Psi_X(x)=  x^{3(\omega+1)},  & \hspace{1cm} &
                \Phi_X(x)=  \omega x^{3(\omega+1)},\\
\Psi_{gCg}(x)= \left[\bar{A}+(1-\bar{A})x^{3(\alpha+1)}
                \right]^{\frac{1}{\alpha+1}},  & \hspace{1cm} &
  \Phi_{gCg}(x)= -\bar{A}\left[\Psi_{gCg}(x)\right]^{-\alpha},\\
\Psi_{q}(x)=  \exp \left[3\omega_1(x-1)\right]x^{3(1+\omega_0-\omega_1)},
   & \hspace{1cm} &
  \Phi_{q}(x)=  \left[\omega_0+\omega_1 (x-1)\right] \Psi_{q}(x),
\end{array}
\end{equation}
where the sub-indexes refer to
each of the state equations: radiation ($r$), matter ($m$),
cosmological constant ($\Lambda$), X-fluid ($X$), Chaplygin gas
($gCg$) and quintessence ($q $).

\begin{figure}[!ht]
\includegraphics[width=.5\linewidth]{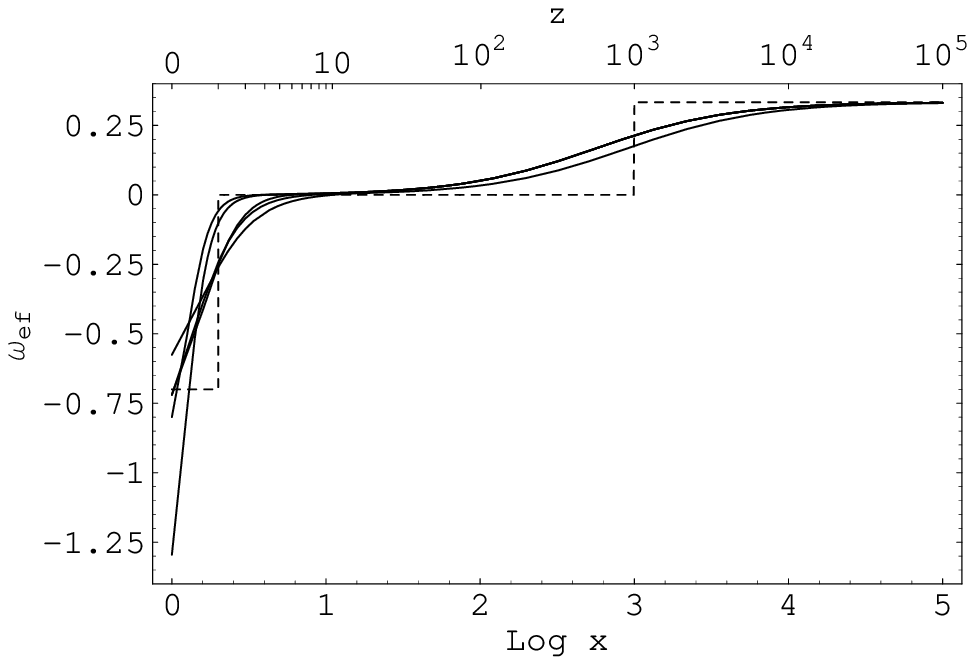}
\includegraphics[width=.5\linewidth]{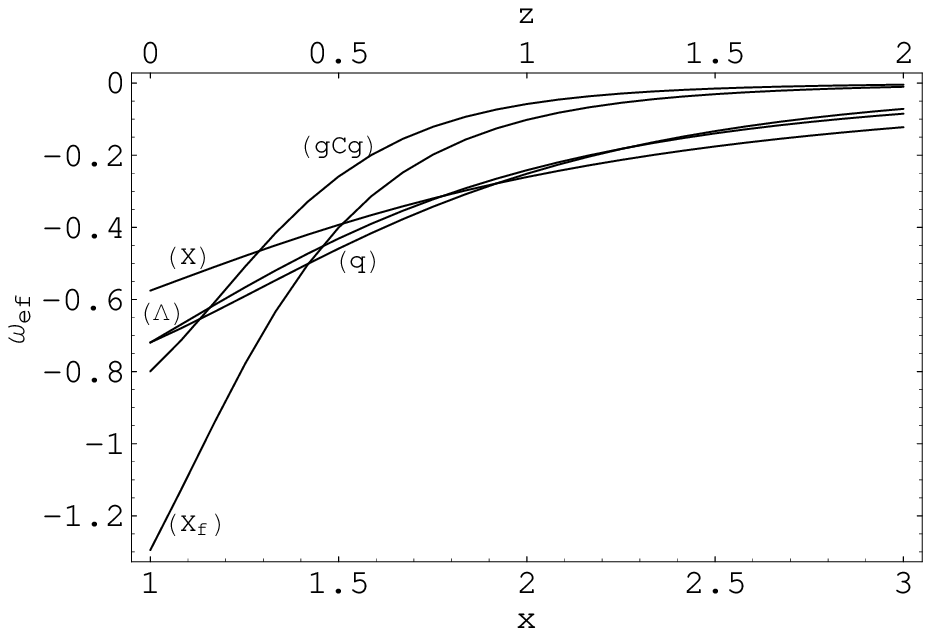}
\caption{Effective equation of state in different cosmologies, all
characterized by three fluids: radiation, matter and dark energy.
The upper panel clearly shows the transition between the radiation-
and matter-dominated phases and sketches the  behavior due to the
dark energy models investigated. The dashed line represents a
model for which the phase transitions are instantaneous and the
dark energy fluid is the cosmological constant. The lower panel
shows in detail the effective equation of state at the low
redshift zone of the upper panel for: cosmological constant
($\Lambda$), X-fluid ($X$), phantom ($X_f$), quintessence ($q$)
and Chaplygin gas ($gCg$). } \label{wef}
\end{figure}

Therefore, the dynamics of the Universe may be computed with no
need of truncating the equation of state  in particular redshifts
between a certain phase and the next one. The effective equation
of state is defined as $p \equiv \omega_{\mathrm{ef}}(x) \rho$,
where
\begin{equation}\label{eosef}
\omega_{\mathrm{ef}}(x) = \frac{3H^2}{8\pi G} \sum_{i} \rho_i(x)\omega_i=
\frac{\sum_i \omega_i \Omega_{i_0}\Psi_i(x)}{\sum_i \Omega_{i_0}\Psi_i(x)}.
\end{equation}
The evolution of $ \omega_{\mathrm{ef}}(x)$ is plotted in \fref{wef}.
The present density parameter of each component is $\Omega_{r_0}=5\times
10^{-4}$; $\Omega_{m_0}=0.28$; and $\Omega_{de_0}=1-\Omega_{m_0}-\Omega_{r_0}$,
except for the Chaplygin gas, which acts as a unified
dark matter-energy fluid: $\Omega_{gCg_0}=1-\Omega_{r_0}$.
The dark energy parameters considered are:
X-fluid: $\omega=-0.8$; phantom: $\omega=-1.8$;
quintessence: $\omega_0=-1,\omega_1=-0.2$; and Chaplygin gas:
$\bar{A}=0.8,\alpha=1$. These values are in agreement with the
observational constraints presently available
(e.g.~see \cite{Lima:04} and references therein).
\Fref{wef} also shows the step-like effective
equation of state (hereafter \emph{step-eos}, dashed line)
which relates  the
second simplifying assumption to be discussed below.

\section{Semi-analytic approach to the problem}\label{semi-analytic approach}

\subsection{The thin-horizon and step-eos assumptions}
\label{assumptions}
\Eref{gw2} has asymptotic solutions which are given by:
\begin{eqnarray}\label{asymptotic1}
h(k,\tau)=  \exp(-ik\tau) \left[a(\tau)\sqrt{2k}\right]^{-1},
\hspace{1cm}  & k^2 \gg |a''/a|
\Leftrightarrow \tau \gg \tau_k %\Leftrightarrow k \gg aH
\\ \label{asymptotic2}
h(k,\tau)=  A_k \ ,
\hspace{1cm}  & k^2 \ll |a''/a|
\Leftrightarrow \tau \ll \tau_k %\Leftrightarrow k \ll aH
\end{eqnarray}
where $A_k $ is a constant and $\tau_k $ is
the transition time when the $k$ mode crosses the horizon ($k^2 = |a''/a|$).
The thin-horizon (or super-adiabatic) approximation ignores the
intermediate cases, $k^2 \sim |a''/a|$,
joining the  solutions \eref{asymptotic1}
and \eref{asymptotic2} at $\tau = \tau_k$, by means of the condition
 $A_k=\left[a(\tau_k)\sqrt{2k}\right]^{-1} \exp (-ik\tau_k)$.

The calculation of cosmic gravitational waves using the thin-horizon approach
is very simple -- it is enough to compute the scale factor  integrating
\eref{friedmann} -- and it becomes still simpler
if the step-eos is assumed. These first calculations were accomplished by
Grishchuk in the early seventies
(see \cite{Grishchuk:01,Grishchuk:04} for review)
considering the radiation and matter phases only and have been recently
extended to include a subsequent $\Lambda$ phase \cite{Zhang:05}.

A formal treatment removing the thin-horizon assumption and taking
an arbitrary number of successive step-eos phases (all with
power law scale factor) was
developed by Maia \cite{Maia:93}, who obtained the formal solution
for \eref{gw2} \cite{Boyle:05}:
\begin{equation} \label{sol instant}
h(k,\tau)=h(k,\tau_i)\Gamma(m+1/2)\left[ k\tau/2 \right]^{1/2-m}
J_{m-1/2}(k\tau), \hspace{.5cm} m=\frac{2}{1+3\omega}
\end{equation}
where $\tau_i$ is an initial time and $k\tau_i \ll 1$ and $h'(k,\tau_i)=0$
were assumed;
$\Gamma(m+1/2)$ and $J_{m-1/2}(k\tau)$ are the Gamma and first type Bessel
functions.  \Eref{sol instant} is
valid for equations of state  $p=\omega \rho$ and does not
include more exotic cases (as the gCg).
This formalism has been recently used
\cite{Zhang:06,Grishchuk:06,Efstathiou:06}
% citar o artigo novo do grishchuk tambem!
to obtain  the gravitational wave spectrum considering the
$\Lambda$CDM phase. However, the step-eos  is still maintained in
these recent works.

\subsection{Beyond the simplifying assumptions}
Using the fairly known relation  $x=1+z=\frac{a_0}{a}$, $a_0=1$, equation
\eref{gw1} may be rewritten (performing a first change of variables from
$t$ to $a$, then substituting \eref{friedmann} into
the resulting equation
and finally applying a second change, from $a$ to $x$) in the
form:
\begin{eqnarray}
\label{gw_DE}
 \frac{d^2{\mathsf h}}{dx^2} + \left(\frac{2}{x}+\frac{3x^2}{2} \frac{A}{B}
 \right)\frac{d{\mathsf h}}{dx} +
\left(\frac{{\mathsf k}^2}{B}-\frac{2}{x^2}+3x
 \frac{A}{B} \right){\mathsf h} = 0 \ , \\
A  \equiv  \sum_{i=m,r,ee}x^{-3}\left[\Psi_i(x)+\Phi_i(x)\right] \Omega_{i_0},
\hspace{1cm} B  \equiv  \sum_{i=m,r,ee} \Psi_i(x) \Omega_{i_0}, \nonumber
\end{eqnarray}
where  ${\mathsf k}$ is the (non-dimensional) wave number:
${\mathsf k} = k/H_0$.
%and the dots correspond to derivatives with respect to $x$.
A convenient new variable $\eta$ may be introduced taking
\begin{equation}
\label{def_mu}
{\mathsf h} x^2  = \eta \ \   \exp\left[\int
\left(x^{-1}-3Ax^2/4B\right)
{\mathrm d}z\right]
\end{equation}
as a definition. The resulting equation is similar to \eref{gw3}:
\begin{eqnarray}
\label{eq_mu}
\frac{d^2\eta}{dx^2}+\left({\mathsf k}^2/{\cal B}+{\cal A}\right)\eta =0,
\hspace{1cm} \mathrm{where}
\\
{\cal A}  \equiv  2/x^2 + f^2 x^4 + \frac{df}{dx} x^2,
\hspace{1cm} {\cal B} \equiv B,
\hspace{1cm} f  \equiv  A/B.  \nonumber
\end{eqnarray}
Therefore, solving  equation \eref{gw_DE} is equivalent
to integrate \eref{eq_mu} using \eref{def_mu} to recover the solution for
${\mathsf h}$.  Formally,  \eref{eq_mu} and \eref{gw3} are similar to each other
and at a first glimpse this new formulation offers no advantage over that
one. However, a second look at \eref{eq_mu} shows a different scenario.

For an effective equation of state $p=\omega_{\mathrm{ef}} \rho$, one has
\begin{equation}
{\mathsf k}^2/{\cal B}+{\cal A}={\mathsf k}^2 x^{3n+3}+(n+2)(n+3)x^{-2},
\hspace{1cm} n=-\omega_{\mathrm{ef}}-2.
\end{equation}
If $n$ does not vary with $x$, then the solution of \eref{eq_mu} is analogous
to \eref{sol instant}:
\begin{eqnarray}\label{sol_analitica}
\eta(x)& = &
\left(\frac{k}{5+3n}\right)^{\frac{1}{5+3n}}x^2 \times \nonumber \\
& & \left[
c_1\Gamma (1-l) J_{-l}\left(\frac{2 k x^{\frac{5+3n}{2}}}{5+3n}\right)+
c_2\Gamma (1+l) J_{l}\left(\frac{2 k  x^{\frac{5+3n}{2}}}{5+3n}\right)
\right],\\
l & = & i\frac{\sqrt{23+20n+4n^2}}{5+3n}. \nonumber
\end{eqnarray}
The assumption $\omega_{\mathrm{ef}}=$ constant is valid for sufficiently small
intervals of $x$ and a feedback process may be used to consider
its variations.
Establishing a set of solutions such  that each
element $\eta_i(x)$ corresponds to an interval $[x_{i},x_{i+1}]$,
the solutions
are all given by \eref{sol_analitica} but with varying values of
$(c_1,c_2)_i$.
A set of constants $(c_1,c_2)_i$
is associated to the solution set.
Taking the initial conditions  from a given inflationary model,
the constants $(c_1,c_2)_0$ of $\eta_0(x)$ are obtained. This first solution
is valid since $x_0=x_{\mathsf k}$, when the  $\mathsf k$ mode
crosses-out the horizon. A recurrence relation schematically expressed as
\begin{equation}
\left.
\begin{array}{c}
\eta_{i}(x_i) =\eta_{i-1}(x_i)  \\
\eta_{i}'(x_i)=\eta_{i-1}'(x_i) \\
x_i=x_{i-1}+\Delta x
\end{array}
\right\}
\Rightarrow (c_1,c_2)_i
\end{equation}
gives the remaining constants and synthesizes the semi-analytic
method here employed.
%(its precision is limited by the size of  $\Delta x$, taken as small as
%necessary).

\subsubsection{Initial conditions} \label{initial cond section}
The necessary initial conditions are defined in the inflationary era
(see e.g. \cite{Lyth:99} for review),
when the super-adiabatic approach is applicable.
During the slow-roll phase one has $\dot{H} \approx 0$ and
in the exact de Sitter case,  $\dot{H}=0$, so that the asymptotic regimes
$k^2 \ll |a''/a|$ and
$k^2 \gg |a''/a|$  are equivalent to $k \ll aH$ and $k \gg aH$, respectively.

Assuming that the perturbations start all inside  the horizon -- with
physical wavelength smaller than the Hubble radius ($k>aH$) --
it is possible to take the solution (\ref{asymptotic1})
for $\tau> \tau_k$. Ignoring the oscillatory part,
the initial condition to be taken is
\begin{equation}\label{cond inic}
h_k(\tau_k) = \frac{1}{a(\tau_k)\sqrt{2k}}=\frac{H}{\sqrt{2}k^{3/2}},
\hspace{1cm} k=a(\tau_k)H,
\end{equation}
where $a(\tau)$ and, consequently, $H$
depend on the inflationary model adopted.
In the  slow-roll regime $H$ is proportional to the  potential
of the inflationary scalar field and for the de Sitter model
this potential is exactly constant so that
\begin{equation}\label{Hout}
H^2=\frac{8\pi}{3 m_{Pl}^2} V(\varphi)=\frac{8\pi}{3}M m_{Pl}^2, \hspace{1cm}
M \equiv \frac{V(\varphi)}{m_{Pl}^4}
\end{equation}
where $M$ is the energy density scale of inflation.
The energy scale $E_{\varphi}$ is the only free parameter
of this model for which $V(\varphi)=E_{\varphi}^4$ and
% The non-dimensional parameter $M$ is related to $E_{\varphi}$ by
$E_{\varphi}=M^{1/4} m_{Pl}$.

Equations \eref{cond inic} and \eref{Hout} lead immediately to the result
\begin{eqnarray}\label{cond inic 2}
 h_k(\tau_k)=
\sqrt{\frac{4\pi M}{3}} \ m_{Pl} \ k^{-3/2} \ \Leftrightarrow \
{\mathsf h}_{\mathsf k}(x_{\mathsf k})=
\sqrt{\frac{4\pi M}{3}} \ m_{Pl} \ {\mathsf k}^{-3/2} \
x^{-2}_{\mathsf k}.
\\\label{cond inic 4}
h_k'(\tau_k)=0 \ \Leftrightarrow \
{\mathsf h}'_{\mathsf k}(x_{\mathsf k})=
-\sqrt{\frac{16\pi M}{3}} \ m_{Pl} \ {\mathsf k}^{-3/2}
\ x^{-3}_{\mathsf k}.
\end{eqnarray}
To conclude the initial conditions set-up
an expression for $x_{\mathsf k}$ is needed, for each mode ${\mathsf k}$
crosses the horizon in a different redshift. That is done using the condition
${\mathsf k}=a(x_{\mathsf k}) H/H_0$, so that
\begin{equation}\label{x inic}
x_{\mathsf k}=\sqrt{\frac{8\pi M}{3}} \ m_{Pl} \ (H_0 {\mathsf k})^{-1}.
\end{equation}
%The system of initial conditions composed by
%\eref{cond inic 2}, \eref{cond inic 4} and \eref{x inic}
%is enough to solve the problem.

\section{Gravitational wave spectrum: results and discussion}
\label{gw spectrum}

\subsection{The power spectrum redefined}
The power spectrum \eref{pwr spectrum t}
is calculated  using the semi-analytic method described above:
\begin{equation}\label{delta h}
\Delta^2_{t}({\mathsf k},x)=\frac{64 \pi G}{2 \pi^2}
(H_0 {\mathsf k})^3 x_{\mathsf k}^4
|x^2{\mathsf h}_{\mathsf k}(x)|^2.
\end{equation}
Knowing that the largest observable wavelength corresponds to the
size of the current horizon, $1/H_0$, the lower limit of physically interesting
frequencies is obtained: ${\mathsf k}=1$ or
$\nu =H_0 \approx 2\times 10^{-18}$Hz (this was the first ${\mathsf k}$
leaving the horizon and did not return  until $z=0$).
%In the super-adiabatic approach its amplitude is exactly equal to the initial value and, even in the (more precise) calculation  presented in this work, it is reasonable to expect that the information about the initial conditions is better preserved for ${\mathsf k} \sim 1$.
The upper limit, in turn, corresponds to the last mode leaving
the horizon at the end of inflation. Considering that the radiation
era starts immediately after inflation (instantaneous and
$100\%$ efficient reheating processes) and that
$\rho_r = \rho_{r_0}x^4$, the end of inflation is:
$x_{\mathrm{end}} = (8\pi M/3)m_{Pl}^2/(H_0^2\Omega_{r_0})$.
Using \eref{x inic}, the maximum wave number is immediately obtained.

Another important quantity is the  amplitude, correlated to
$\Delta^2_{t}({\mathsf k},x)$ through:
\begin{equation}\label{hc}
{\mathsf h}_c({\mathsf k},x)=\sqrt{\frac{\Delta^2_{t}({\mathsf k},x)}{2}}
\end{equation}
and  of great interest when the detectability of
primordial gravitational waves is computed. The  amplitude
is approximately equal to the variation observed in the distance $L$
between two proof masses submitted to a gravitational radiation flux:
${\mathsf h}_c({\mathsf k},x) \approx \Delta L/L$.

The third quantity of interest is the spectral energy density,
defined as
\begin{equation}
\label{omega gw}
\Omega_{gw}(k,\tau) \equiv \frac{3H^2}{8\pi G}
\frac{d\langle0|\hat{\rho}_{gw}|0\rangle}{d \ln k}, \hspace{1cm}
\rho_{gw}=-T^0_0.
\end{equation}
To calculate $\Omega_{gw}(k,\tau)$ the
action \eref{action2} is considered to  obtain the stress-energy tensor
%(actually, only the 00 component is needed)
%\begin{equation}
%T_{\mu \nu}=-2\frac{\delta {\cal L}} {\delta g^{(B)\mu \nu}} +
%g^{(B)} _{\mu \nu}{\cal L}.
%\end{equation}
%so that the classical energy density associated to the gravitational waves is
%\begin{equation}
%\rho_{gw}=-T^0_0=\frac{1}{64\pi G}\frac{(h'_{ij}) ^2+(\nabla h_{ij}^2)}{a^2}.
%\end{equation}
and the same procedure that led to \eref{pwr spct t def} can be applied
to obtain,
\begin{eqnarray}
\langle0|\hat{\rho}_{gw}|0\rangle=\int^{\infty}_0 \frac{k^3}{2\pi^2}
\frac{(h'_{ij})^2+k^2 h_{ij}^2}{a^2}\frac{dk}{k}, \nonumber\\
\label{omega gw2}
\Omega_{gw}(k,\tau)=\frac{8\pi G}{3H^2}\frac{k^3}{2\pi^2}
\frac{(h'_{ij})^2+k^2 h_{ij}^2}{a^2} = \frac{1}{12}
\frac{k^2\Delta^2_h(k,\tau)}{a^2H^2},
\end{eqnarray}
where in the last term of \eref{omega gw2} it was considered that
$|h'(k,\tau)|^2=k^2|h(k,\tau)|^2 $.
As a function of $\mathsf k $ and $x $, the spectral energy density is
\begin{equation}\label{omega gw3}
\Omega_{gw}({\mathsf k},x)=\frac{x^2}{12}
\frac{(H_0{\mathsf k})^2\Delta^2_{t}({\mathsf k},x)}{H^2(x)}.
\end{equation}

The spectra thus defined
are all interesting. They contain  information
on the amplitude of the tensor perturbations \eref{delta h},
their detectability  \eref{hc} and the energy density of each
 mode \eref{omega gw3}.

\subsection{During the inflationary era}\label{inflation}
\begin{figure}[!ht]
\includegraphics[width=.7\linewidth]{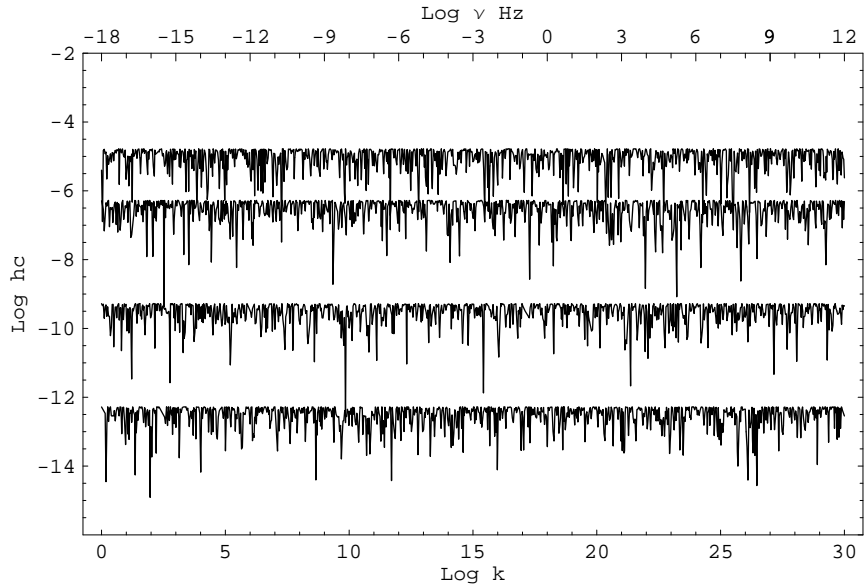}
\caption{Amplitude spectrum of relic gravitational waves taken
immediately after inflation for four energy scales: $M=10^{-14}$, $10^{-16}$,
$10^{-20}$, $10^{-24}$ (from top to bottom).
}
\label{infM}
\end{figure}

%During inflation, the modes leave the horizon in the inverse order of
%their wavelengths and, as it was argued before,
%at redshift $x_{\mathrm{fi}}$, when inflation ends,
%the modes of interest are those which left the horizon.
According to the initial conditions \eref{x inic} and
\eref{delta h}, the amplitude spectrum
at $x=x_{\mathrm{end}}$ is flat,  as shown in \fref{infM}
for four energy scales:
$M=10^{-24}$ $(E_{\mathrm{inf}} =10^{-6}m_{Pl}) $;
$M=10^{-20}$ $(E_{\mathrm{inf}} =10^{-5}m_{Pl}) $;
$M=10^{-16}$ $(E_{\mathrm{inf}} =10^{-4}m_{Pl}) $;
$M=10^{-14}$ $(E_{\mathrm{inf}} =10^{-3.5}m_{Pl}) $.
The spectra were calculated at $x_{\mathrm{end}} (M) \approx 10^{25}-10^{28}$.

Defining $\delta_{\mathsf h} \equiv
\Delta \log {\mathsf h}_c / \log {\mathsf h}_c $
and $\delta_M \equiv \Delta \log M / \log M $, one notices that
$\delta_{\mathsf h} \approx 3/4 \delta_M $.
For $x <x_{\mathrm{end}} $, the spectrum evolves such that
the amplitude of modes inside the horizon is always decreasing
(sections \ref{decelerated} and \ref{accelerated}), while the external
ones remain almost constant.
This fact allows, in principle,
to establish an upper limit for $M$. Since the metric
perturbations leave their imprint on the CMB anisotropy spectrum,
the maximum  amplitude should not be larger than
$\Delta T/T \approx 10^{-5}$  at $x \sim 10^3-10^4$, when  photons
decouple. This  constraint implies that $M \leq 10^{-14}$
(this conservative estimate is discussed in \sref{cmb}).

A remarkable characteristic to be noticed in \fref{infM} and the following ones
is the oscillatory shape of the spectra. Far from being a purely numeric
effect, those oscillations result form the very mathematical solution of
\eref{gw2}:
% . One can see, from the analytic solution explicitly shown in \eref{}, that
there are oscillations not only in time, but also in the Fourier
space.

\subsection{During the non-accelerated expansion} \label{decelerated}
\begin{figure}[!ht]
\includegraphics[width=.7\linewidth]{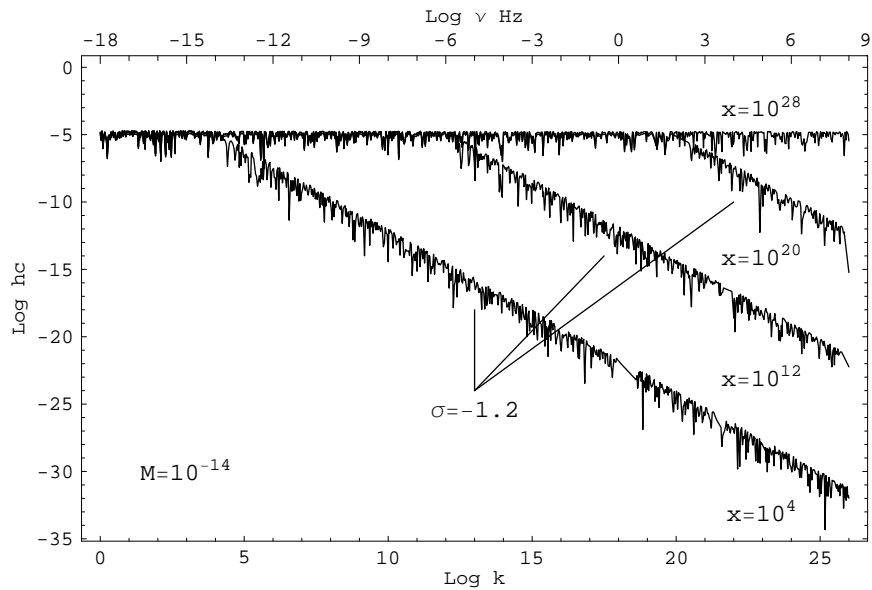}
\caption{Tensor power spectrum (chirp amplitude), for  $M=10^{-14}$, at
different redshifts inside the radiation-dominated phase. The slope
of the spectrum is $\sigma \approx -1.2$, which is comparable to previous
estimates \cite{Zhang:05}.
}
\label{rad14}
\end{figure}

%The spectrum of \fref{infM} is the initial condition for the other phases.
The result of the semi-analytic procedure is shown in
\fref{rad14}, for different redshifts inside the radiation era.
The modes of smaller wavelength cross-back the horizon at larger redshifts
so that the modifications in the power spectrum start
at the higher frequencies and  the information about the initial
conditions are better preserved for low frequencies. In the
area corresponding to the modes that already crossed the horizon, the
slope is $\sigma \approx -1.2$.

The transition from radiation- to matter-dominated era begins at about
$x \approx 10^{4}$ and finishes at $x \approx 10^{2}$ (\fref{wef}).
It is important to highlight that the shape of the spectrum
carries information mainly about: (1) the initial conditions
determined in in the primordial Universe and (2)
the evolution of the effective equation of state -- or equivalently
of the scale factor. While out of the horizon each
${\mathsf k}$ is maintained (almost) unaffected preserving its initial value;
when returning to the horizon, it begins to evolve; however,
when $1/H$  is much larger than ${\mathsf k}$, \eref{sol_analitica}
tends to the asymptotic regime \eref{asymptotic2}, when all of
the modes fall proportionally to $a^{-1}$. For this reason, one observes
that the high frequency region in \fref{rad14} maintains
the same slope (only the amplitude varies) during the
subsequent phases.

\begin{figure}[!ht]
\includegraphics[width=.7\linewidth]{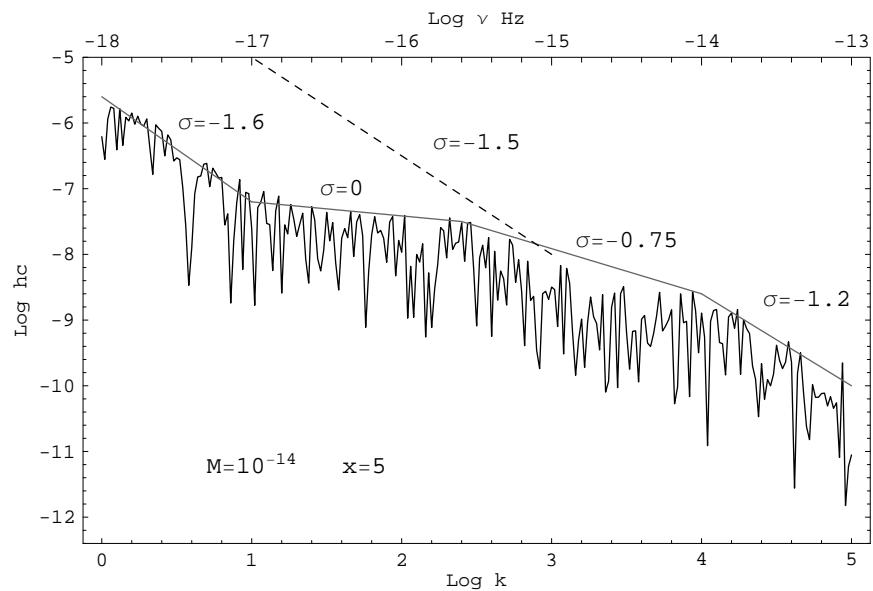}
\caption{Low frequency region of the amplitude spectrum of
primordial gravitational waves, calculated at $x=5$, for
$M=10^{-14}$. The dashed line corresponds to the result expected
from the step-eos assumption. } \label{mat14b}
\end{figure}

\Fref{mat14b} shows in detail the low frequency region of the power spectrum
calculated at $x=5$, far inside the matter-dominated era, when almost the
whole spectrum has been  modified from its initial
condition.
The spectrum has an almost null slope in the region $10^1 \lesssim
{\mathsf k} \lesssim 10^{2.4}$ while in $10^{2.4} \lesssim
{\mathsf k} \lesssim 10^4 $ its inclination is modified because of
the variation of $\omega_{\mathrm{ef}}$. That feature is
completely different of what is obtained from an step-eos.
Modelling $\omega_{\mathrm{ef}}$ with a step-function (dashed line
in \fref{wef}), other authors
\cite{Grishchuk:01,Zhang:05,Boyle:05,Zhang:06} obtain a spectrum
with  $\sigma \approx -1.5$ for ${\mathsf k} <10^3 $
\cite{Zhang:06}. For comparison a dashed line is included in
\fref{mat14b}, corresponding to the slope obtained with the
step-eos.

A last comment is  concerned with the region $1 <{\mathsf k} \lesssim 10 $ of
\fref{mat14b}. The picture was taken in $x=5$ and so the modes within that
region are in the threshold of the transition from ``outside'' to
``inside''  the horizon (${\mathsf k} \sim aH$). Since  the  thin-horizon
simplification is not considered in this work,
that region shows a transitory slope
($\sigma \approx -1.6 $), that disappears for $x$ smaller than 5, as the
modes enter the horizon. This characteristic is not perceptible
in the previous illustration due to the scale.

\subsection{During the accelerated expansion}\label{accelerated}
\begin{figure}[!ht]
\includegraphics[width=.7\linewidth]{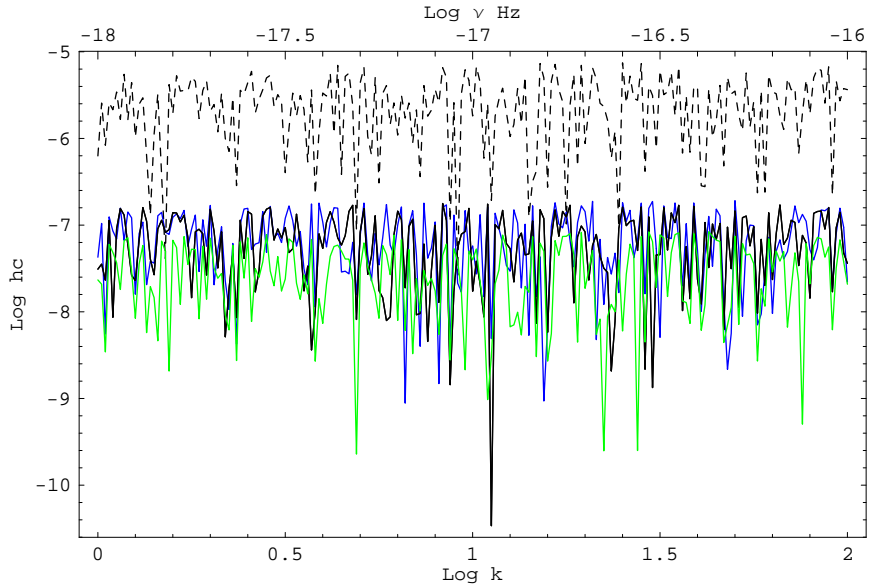}
\caption{Present spectrum ($x=1$) of gravitational waves for $M=10^{-14}$.
The dark energy models considered are: cosmological constant (solid),
Chaplygin gas (blue), X-fluid (green), phantom (dashed), and quintessence
(gray). The parameters used are the same as in \fref{wef}.
}
\label{ee14a}
\end{figure}

Nearly all the modes are already inside the horizon when the accelerated
expansion begins. For that reason, the shape of the
spectrum remains constant and only the amplitude is altered in that phase.
\Fref{ee14a} shows a narrow slice of the current spectrum,
so that the effect of the different models of dark energy can be better observed.

The first characteristic to stress with regard to \fref{ee14a} is
the complete degeneracy between cosmological constant and
quintessence models ($\omega_0=-1,\omega_1=-0.2$), whose curves
overlap completely. Moreover, the Chaplygin gas ($\bar{A}=0.8$ and
$\alpha=1$) is also almost degenerate with those two curves.
Observing \fref{wef} it is easy to notice that these three cases
have very similar behavior and this degeneracy is expected. The
X-fluid ($\omega=-0.8$), in turn, leads to an amplitude smaller
than the cosmological constant while the phantom fluid
($\omega=-1.8$) acquires a larger amplitude. The difference is
greater for phantom than non-phantom fluid  because the chosen
parameters are such that
$|\omega_{X_f}-\omega_{\Lambda}|>|\omega_{X}-\omega_{\Lambda} | $.
Since the dark energy begins to act very recently $x \sim 2$, the
signature of  different models is very weak -- to produce a
difference of $\approx 2 $ in $\log {\mathsf h}_c $ scale, it is
necessary to vary $\omega$
 of
almost one unit (the more negative is the equation of state, the larger
is the amplitude).

Although just  a particular choice of parameters is shown, other
possible combinations have been tested.
The  results allow to state that:
(1) for gCg \cite{Fabris:04,Soares-Santos:05}
the parameter $\alpha$  does not have any visible effect, while $\bar{A} $
increases  the amplitude as it varies from $0$ (corresponding to the minimum
stipulated by the spectrum without dark energy) to $1$ (reaching the
$\Lambda$CDM curve);
(2) the X-fluid  \cite{Soares-Santos:05a} has only one parameter, $\omega$,
that plays the same role of $\bar{A} $, but it assumes values more
negatives  than $-1$ and, in that case, the generated spectrum is above
$\Lambda$CDM;
(3) quintessence, in general, is completely degenerate with the X-fluid,
unless that  $\omega_1 \sim 1$ (or greater) is assumed,
but that does not have any physical sense for in this case the
quintessence  would affect the dynamics of the early  Universe, being
in contradiction with the most accepted cosmological models.

%Concluded the analysis of the spectra,
%it remains to investigate the detectability of those
%primordial gravitational waves and this is the objective of the next section.

\section{Search for empirical counterparts}

%Once obtained the theoretical spectrum of cosmic gravitational waves, the natural question concerns the empiric counterparts of those forecasts.

\subsection{Direct attempts}\label{direct}
The term ``direct detection'' should be strictly used only for experiments
whose principle  is to measure the deformation of the space-time submitted to
tensor perturbations. The interferometric experiments lay in that class.
However, it is common to include under that label detectors that measure the
deformation of a proof mass in resonance and that is done in this work.
The quantities $\Delta^2_t $,
$\Omega_{gw} $ and ${\mathsf h}_c $  characterize, in an equivalent
way, the gravitational wave background and do not depend on the
detector. However, the signal  $S(t)$ registered by an hypothetical detector
will have a component $s(t)$, representing the observable, and another,
the noise $r(t) $, which depends on the characteristics of the detector:
$S(t)=s(t)+r(t) $.
The mean (square) contribution of these components are \cite{Maggiore:00}:
\begin{equation}\label{mean sqrt}
\langle s^2(t) \rangle = \frac{F}{2} \int^{\infty}_0
\nu^{-1} {\mathsf h}_c^2 d\nu, \hspace{1cm}
\langle r^2(t) \rangle =  \int^{\infty}_0 h_r^2 d\nu,
\end{equation}
where $F$ is an efficiency factor representing the loss of
sensitivity due to the fact that gravitational waves come from all
directions, while the detector is only maximally sensitive in some
preferential ones. That expression for $\langle s^2(t)\rangle$ is
valid for a stochastic background and the gravitational radiation
may be considered detectable if
\begin{equation}\label{detect condition}
{\mathsf h}_c > \sqrt{\frac{2 \nu}{F}}h_r.
\end{equation}
%The detectors sensitivity is usually
%expressed in terms of $h_r $ (dimension of Hz$^{-1/2} $).

\subsubsection{Resonant masses}
Schematically, a resonant mass
detector  is a  (cylindrical or spherical) massive solid
mass whose mechanical oscillations
are excited by gravitational radiation and converted to electric
sign by a sensor.  The variation
of the mass length is a sum of all its vibration modes, but the
sensor is filtered to receive only the fundamental frequency, being
therefore a narrowband detector.

In the case of a \emph{cylindrical bar}, the
efficiency factor is $F=8/15$, for the preferential direction of
incidence is any direction perpendicular to the bar (if the sensor is located
at one of its ends).
The sensibility of the last generation experiments is
\cite{Coccia:00}
$h_r \approx 10^{-22} \mathrm{Hz}^{-1/2} $,
at $\sim 900$Hz. Using \eref{detect condition}
the  minimum detectable amplitude of those experiments is:
\begin{equation}\label{hc min bars}
{\mathsf h}_c \gtrsim 5,8 \times 10^{-21},
\hspace{1cm} {\mathrm{for}} \ \ {\mathsf k} \approx 4
\times 10^{20}.
\end{equation}

A \emph{spherical detector} has a larger mass for the same resonance frequency
implying in a greater cross-section. Besides, it is sensitive to different
directions and polarizations of the incident  radiation.
The sensibility of those
detectors is planned to be \cite{Aguiar:05}
$h_r \approx 10^{-23} \mathrm{Hz}^{-1/2}$, for
frequencies between $200$Hz and $2$kHz.
\Eref{detect condition} here implies that resonant spheres are able to detect
gravitational waves of up to
\begin{equation}\label{hc min spheres}
{\mathsf h}_c \gtrsim 2,7 \times 10^{-22},
\hspace{1cm} \mathrm{for} \ \ {\mathsf k} \approx 1
\times 10^{20}.
\end{equation}

Comparing  bars \eref{hc min bars} and spheres
\eref{hc min spheres}, one notices an improvement of one order of  magnitude,
but that  is still not enough to detect waves of
cosmological origin.

\subsubsection{Interferometers}
\label{interfereometers}
Laser interferometry is used in large detectors to measure the displacement of
free-falling masses submitted to gravitational radiation.
The basic components of those detectors  are two arms of length $L$ connected
at one end and three masses (one at the junction point, two at the free ends),
constituting three pendulums.
The preferential incident  direction is
perpendicular to the plan of the arms %and, in this situation,
%the free-falling masses at one arm will be moved away of a distance
%$\Delta L=\frac{1}{2}L{\mathsf h}_c $ while the masses of the other arm
%will approximate of the same factor
%(for polarization $+ $ and perpendicular arms):
and $F=2/5 \sin^2 \theta $, where $\theta$ is the angle formed by the arms
\cite{Maggiore:00}.

Unlike resonant masses, interferometers are broadband apparatus whose
maximum wavelength is limited by the size of the arms.
The main interferometric detectors are:
LIGO \cite{Gustafson:99}, with a 4km arm length; VIRGO
\cite{Acernese:05a,Acernese:05}, 3km-sized;
GEO600 \cite{Grote:05} and TAMA300 \cite{Ando:99}, with
600 and 300 meters, respectively; AIGO/ACIGA \cite{McClelland:00}, planned
to have 4km; and
finally, the most ambitious projects, BBO \cite{Corbin:06} and
LISA \cite{Danzmann:03} are  space interferometers where three arms
($5 \times 10^6 $km in LISA and $5 \times 10^5 $km in BBO)
are arranged in an equilateral triangle.

\begin{figure}[!h]
\includegraphics[width=.7\linewidth]{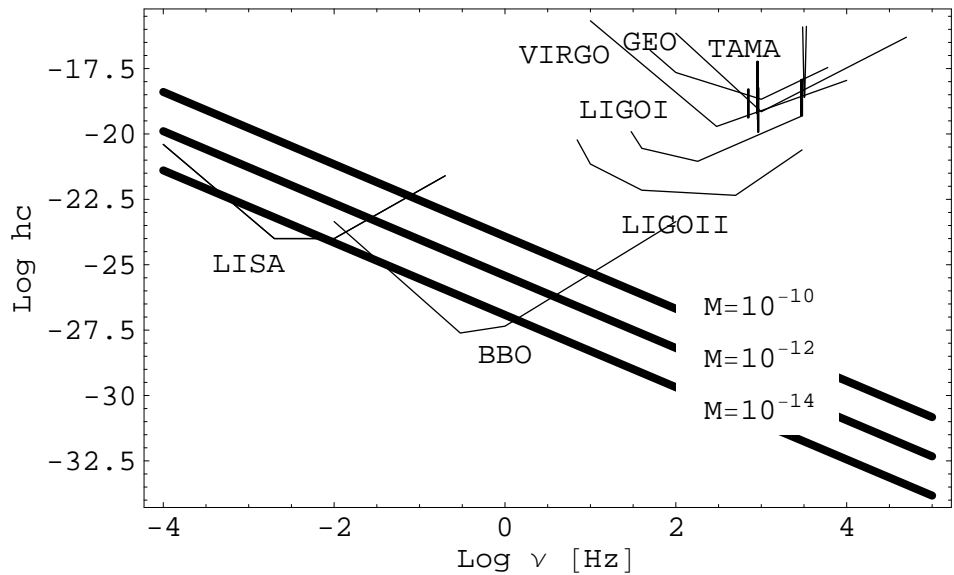}
\caption{Gravitational wave spectrum for three inflationary energy scales
($M=10^{-14}$, $10^{-12}$, $10^{-10}$, from bottom to top, thick lines) in
comparison with sensitive estimates for the detectors.
}
\label{hcdet}
\end{figure}

The detectability estimate is performed using \eref{detect condition} once
more. The results are shown in  \fref{hcdet}.
In the high frequency band,  resonant masses and ground-based
interferometers are shown, all with insufficient sensibility to detect
primordial gravitational waves.
The sensitivity curves of  resonant masses are nearly  reduced to a point
in that scale, for these are, by construction, narrowband detectors.
The interferometric detectors are also
identified in the figure. %From top to
%bottom, right to left, one has TAMA, GEO, VIRGO, LIGO I, LIGO II,
%BBO and LISA.
Only LISA and BBO would be able to detect a cosmological
signal, but both are still in project \cite{Corbin:06,Danzmann:03}.
%, it can be said that the
%detection perspectives, at least with direct atempts, are not immediate.

\subsection{CMB: an indirect attempt}\label{cmb}
In simple terms, one can say that  \cite{Hu:96}
until $z_* \approx 10^3$, the temperature of the Universe is
high enough to ionize the hydrogen and, via Compton scattering,
photons and electrons are coupled; in addition, electrons are
also coupled to barions through electromagnetic interaction. The radiation
pressure offers resistance to  gravitational forces and acoustic
oscillations arise in the plasma. In $z_*$, the
hydrogen recombines (and the photons last scatter); the Universe
becomes then transparent to  photons and compression and rarefaction areas
of the plasma at that redshift represent hot and cold areas, respectively;
besides,  photons suffer gravitational redshift when leaving the potential
wells of the last scattering surface (Sachs-Wolfe effect
\cite{SachsWolfe:67}).
The resulting fluctuations appear as (primary) anisotropies
in the sky. %There is a critical
%wavelength, $\lambda_{{\mathrm{crit}}} = \int c_s d\tau $, corresponding
% to an angular scale $\theta_ * = \sqrt{2 \Omega_0 (1100/z_ *)} \simeq 1^{o}$
%\cite[Cap.~9]{AW:KT90} ,
%above which the
%oscillations do not  establish and the anisotropies do not evolve,
%except by gravitational redshift.

The theoretical calculation of the CMB anisotropies is based on the linear
theory of cosmological perturbations \cite{Giovannini:05}.
The temperature anisotropy
at position ${\bf x}$ and direction ${\bf n} $,
$\Delta_T({\bf x},{\bf n}) $, depends, in principle, on both
the direction and the frequency, but the distortions in frequency are
of second order. A Fourier expansion of $\Delta_T({\bf x},{\bf n}) $,
results in modes that propagate independently of each other.
Assuming  axial symmetry around ${\bf k}$, a
Legendre expansion can also be done:
\begin{equation}\label{Legendre expansion T}
\Delta_T({\bf k},{\bf n})=\sum_l(2l+1)(-i)^l\Delta_{T_l}P_l(\mu),
\hspace{1cm} \mu={\bf k} \cdot {\bf n}/k,
\end{equation}
where $P_l(\mu)$  is the Legendre polynomial of order $l $, $\Delta_{T_l}
$ is the associated multipole moment %, $\mu $ is the cosine of the angle
%between ${\bf k} $ and the  line of sight
and %, to simplify
%the notation,
%$\tilde{\Delta}_T({\bf k},{\bf n}) =
$\Delta_T({\bf k},{\bf n})$ is the Fourier transform of
$\Delta_T({\bf x},{\bf n}) $.  A similar
expression can be written for  polarization
anisotropies $\Delta_P({\bf k},{\bf n})$.

%The anisotropies evolution  is governed by the
The Boltzmann equation, which describes the temporal evolution of
the Stokes parameters of the radiation field, is constituted by a
collisional (referring to the Thompson scattering) and a
non-collisional term. In the case of tensor perturbations, the
resulting equation is \cite{Crittenden:93}
\begin{eqnarray}
{\Delta'}_T^{(t)}+ik\mu \Delta_T^{(t)}  =
-h'-\kappa'
\left(\Delta_T^{(t)} - {\bf Y} \right) \nonumber \\
\label{Boltzmann eqs tensorial}
{\Delta'}_P^{(t)}+ik\mu \Delta_P^{(t)}  =  -\kappa'
\left(\Delta_P^{(t)} + {\bf Y} \right)\\
{\bf Y}  =
\left[\frac{1}{10} \Delta_{T_0}^{(t)}+\frac{1}{35}\Delta_{T_2}^{(t)}+
\frac{1}{210}\Delta_{T_4}^{(t)}-\frac{3}{5}\Delta_{P_0}^{(t)}+
\frac{6}{35}\Delta_{P_2}^{(t)}-\frac{1}{210}\Delta_{P_4}^{(t)}
\right] \nonumber
\end{eqnarray}
where $h$ is governed by equation \eref{gw2} and
\begin{equation}\label{profundidade optica}
\kappa=\int^{\tau_0}_{\tau} \kappa' d\tau
%\hspace{1cm}
%{\cal K} =\kappa' \exp (-\kappa).
\end{equation}
represents the optical depth %and visibility function
(and $\kappa' \equiv a n_e x_e \sigma_T$).
The solutions of \eref{Boltzmann eqs tensorial} may be written in the form
\cite{Seljak:96}
\begin{equation}\label{sol Boltzmann generica 2}
{\cal F}(k,\tau_0)=
\int^{\tau_0}_0 e^{ik\mu (\tau -\tau_0)} {\cal S}(k,\tau) d\tau.
\end{equation}
where ${\cal F}$ is  the  (temperature or polarization) anisotropy,
$\tau_0$ is the present time and
${\cal S}$ is the  corresponding
source term:
\begin{equation}\label{termos fonte gw}
{\cal S}_T^{(t)}=-e^{-\kappa}h'+{\cal K}{\bf Y} \hspace{1cm}
{\cal S}_P^{(t)}=-{\cal K}{\bf Y},
\end{equation}
where ${\cal K} =\kappa' \exp (-\kappa)$ is the visibility function.
Expanding  ${\cal F}$ in terms of tensor multipoles one has
\begin{equation}
\label{multipolos generica}
{\cal F}_l(k,\tau_0)  =  \int^{\tau_0}_0
\sqrt{\frac{(l+2)!}{2(l-2)!}} \ [k(\tau - \tau_0)]^{-2} \
j_l[k(\tau-\tau_0)]
{\cal S}(k,\tau) d\tau \ .
\end{equation}
If the observation takes place at ${\bf x}=0 $,
the expansion of ${\cal F}$ in spherical harmonics is
\begin{equation}\label{expansao harm esf}
{\cal F}=
\sum_{l=0}^{\infty} \sum_{m=-l}^{l} a_{lm}Y_{lm}(\theta,\phi)=
\frac{1}{(2\pi)^{3/2}}\int  d{\bf k}  {\cal F}_l \ ,
\end{equation}
where $Y_{lm} $ is the angular part of the spherical harmonic eigenfunctions,
such that:
\begin{equation}\label{Pl Ylm}
P_l(\mu)=\frac{4\pi}{2l+1}\sum_{m=-l}^{l}Y_{lm}^{*}(\theta,\phi)
Y_{lm}(\theta,\phi), \hspace{1cm} \mu=\cos \theta
\end{equation}
>From \eref{Pl Ylm} and \eref{expansao harm esf}
it results
\begin{equation}\label{alm}
a_{lm}=\frac{(4\pi)(-i)^l }{(2\pi)^{3/2}}
\int d{\bf k} \ {\cal F}_l \ Y^{*}_{lm}(\theta,\phi),
\hspace{1cm}
\langle a_{lm}a_{l'm'}^{*}\rangle \equiv C_l\delta_{ll'}\delta_{mm'} \ .
\end{equation}

The evolution of the Boltzmann equations
does not depend on ${\bf k} $, so it is possible to take
\cite{Ma:95},
${\cal F}_l \rightarrow h_{\mathrm{ini}} {\cal F}_l $,
where $h_{\mathrm{ini}}$ is the initial condition
of the perturbation.
The correlation function is
$\langle h_{\mathrm{ini}}({\bf k})h_{\mathrm{ini}}^{*}({\bf k}')\rangle=
\Delta_t^2(k) \delta({\bf k}-{\bf k}')$
and so the $C_l $ coefficients are determined:
\begin{equation}\label{Cl 2}
C_l^{(t)}  =  \left[\frac{4\pi}{(2\pi)^{3/2}}\right]^2
\int \Delta^2_{t} |{\cal F}_l|^2
\langle Y_{lm}^{*}Y_{lm} \rangle d{\bf k}
 = \frac{2}{\pi} \int \Delta^2_{t} |{\cal F}_l|^2
 k^2 d k
\end{equation}

In short, to calculate the induced tensor perturbations of the
temperature (polarization) $C_l$ spectrum  it is necessary to
compute the multipoles $\Delta_{T_l}^{(t)} $ ($\Delta_{P_l}^{(t)}
$) using \eref{multipolos generica} and the source term ${\cal
S}_T^{(t)} $ (${\cal S}_P^{(t)} $); the result is applied directly
on \eref{Cl 2}, using  the initial power spectrum $\Delta_t^2 $,
given by equation (\ref{pwr spectrum t}) and shown in \fref{infM}.
All information on the evolution of gravitational waves after
inflation is contained in
%$\Delta_{T_l}^{(t)} $ ($\Delta_{P_l}^{(t)} $), i.e., in
the source terms.

From \eref{termos fonte gw} it is noticed that the polarization
source term is suppressed if the media is optically thin, so that
after reionization the  polarization due to gravitational waves is
negligible and the polarization spectrum may be used to verify the
initial conditions obtained in \sref{initial cond section}.
Although both tensor and scalar perturbations affect the CMB, it
is possible to find a mode ($B$ polarization mode) which is
induced only by gravitational waves and this is a promising future
perspective to constraint the initial conditions of the problem.

The temperature source term in \eref{termos fonte gw} is reduced to
${\cal S}_T^{(t)} \approx - h' $ after recombination and,
therefore, it contains information
about the evolution of the power spectrum since then.
Moreover, after
the condition $k \ll aH$ is achieved the amplitude $h $ falls as $a^{-1}$
and  only the modes $k \lesssim k_ {*} $ (where $k_{*}$ is the
wavelength of the order of the Hubble  radius  at recombination time)
contribute to the source term ${\cal S}_T^{(t)} $.

%  You gave graphs is possible to notice that, of the thermodynamic
%point of view, the transition of phase matter-radiation is almost
%instantaneous. The form of the function visibility reinforces the argument
%discussed above that the polarization spectrum only contains information on
%the
%primordial gravitational waves in $x \ simeq 1300 $.

The power spectrum of  temperature anisotropies was calculated
according to the prescription above, using the profiles of $\kappa
$ and ${\cal K} $ obtained from the Saha equilibrium equation and
the envelope of the spectrum corresponding to the $\Lambda$CDM
model with $M = 10^{-14}$. The integration limits
($k_{\mathrm{min}} =1$ and $k_{\mathrm{max}} =1500 \simeq k_{*} $,
$x_{\mathrm{min}} =1 $ and $x_{\mathrm{max}} =1500 $) were chosen
in order to reduce the computacional cost without losing relevant
information. Only the multipoles $l \leq 10 $ were calculated and
the result is a plane spectrum, with amplitude $C_l
l(l+1)/2\pi=41.1 (\mu \mathrm{K})^2 $. The form of the spectrum is
compatible with other theoretical predictions
\cite{Crittenden:93,Efstathiou:06}, as well as with observational
results \cite{Spergel:06}, but the value found is $\sim 25 $ times
lower. This means that the inflationary energy scale $M $ could be
larger than the conservative limit adopted along this work. With
$M = 10^{-10} $, this difference
 is suppressed: $C_l l(l+1)/2\pi=1.01 \times 10^{3}(\mu \mathrm{K})^2
$. The current spectrum of gravitational waves can have in this case the
amplitude given by the superior curve of \fref{hcdet}.

The limit $M=10^{-14} $($E_{\mathrm{inf}} =10^{-3,5}m_{Pl} $) is a
direct consequence  of the condition that the amplitude of tensor
perturbations  should not be larger than the observed CMB
anisotropies and, at a first glimpse it may seem in contradiction
with the result presented above. Under a more careful glance,
however, one notices that this last result is obtained when
considering the physical processes involved in the generation of
anisotropies induced by gravitational waves. Additional physical
ingredients (specifically the ones regarding to the recombination
process) that determine the efficiency with which the
perturbations imprint themselves on the CMB spectrum had to be
considered.  In order to avoid the additional uncertainties
introduced by these new ingredients, the authors have chosen the
most conservative limit, $M=10^{-14} $, when performing the
analyzes of \sref{inflation}. However, other possible values have
been considered in the literature between $M \simeq 10^{-8} $
\cite{AW:KT90} and $M \simeq 10^{-11} $ \cite{Efstathiou:06},
which are consistent with the results here presented.

\section{Summary and conclusions}
The origin and evolution of the primordial gravitational wave
background was computed for different cosmologies. Temporal
evolution of these relics is governed by an oscillator-like
equation with an exciting potential $a''/a$. The initial
conditions are set-up during inflation and, for a de Sitter model,
the initial spectrum is flat. As the modes cross back to the
horizon, ever since the end of inflation until $x \sim 10^{4} $,
the spectrum assumes a slope ($\sigma \approx -1.2 $) in  the high
frequency region (${\mathsf k}>10^{4}$). That slope is maintained
in the subsequent eras. The modes with ${\mathsf k} $ between
$\sim 10^{4} $ and $\sim 10^{2} $ ($\sigma=-0.75$) enter the
horizon during the radiation-matter transition and the step-eos
assumption is shown to be unsupported in these redshifts. The
final spectrum is flat in the low frequency band $1 \leq {\mathsf
k} \lesssim 100$.

The analysis here accomplished also shows that gravitational waves
are not strongly affected by dark energy. Establishing constraints
on dark energy using gravitational waves is not a promising task
(even though one could arrive to the very opposite conclusion
naïvely considering that both of them interact only via
gravitation). One must measure the gravitational wave spectrum
with precision better than $\delta_{{\log}_{10}{\mathsf
h}_c}\lesssim 0,4 \% $ to constrain  $\omega $ within
$\delta_{\omega}<1 \%$ (with the energy scale $M$ independently
fixed and an X-fluid equation of state assumed \emph{a priori}).

Once obtained the theoretical spectrum of cosmic gravitational waves,
the natural question concerns the empiric counterparts of those forecasts.
None among all the already built experiments is sensitive enough to
detect  primordial gravitational waves, because at their operational
frequencies that signal is orders of magnitude below their noise levels.
The perspectives are promising for the future space detectors though.

CMB, in turn,  supplies a concrete possibility of obtaining this
information indirectly, but the  influence of gravitational waves
on the CMB must be understood in detail in order that one can be
able to extract  information from the angular spectra whose
detection is expected to be greatly improved in the near future.
The calculation here performed  establishes a limit on the energy
scale of inflation: $M \lesssim 10^{-10}$.

%%%%%%%%%%%%%%%%%%%%%%%%%%%%%%%%%%%%%%%%%%%%%%%%
%% BACKMATTER
%%%%%%%%%%%%%%%%%%%%%%%%%%%%%%%%%%%%%%%%%%%%%%%%

% acknowledgements
\vspace{.5cm}
This work was financially supported by FAPESP and CNPq.

\bibliography{bib}

\end{document}